\documentclass[aps,preprint]{revtex4}%
\usepackage{amsfonts}
\usepackage{amsmath}
\usepackage{amssymb}
\usepackage{graphicx}%
\begin{document}
\preprint{ }
\title[$T$-matrix]{Calculation of the two-body $T$-matrix in configuration space.}
\author{George Rawitscher}
\affiliation{Department of Physics, University of Connecticut, Storrs, CT 06268}
\pacs{PACS number}

\begin{abstract}
A spectral integral method ($IEM$) for solving the two-body Schr\"{o}dinger
equation in configuration space is generalized to the calculation of the
corresponding $T-$matrix. It is found that the desirable features of the
$IEM$, such as the economy of mesh-points for a given required accuracy, are
carried over also to the solution of the $T-$matrix. However the algorithm is
considerably more complex, because the $T-$matrix is a function of two
variables $r$ and $r^{\prime}$, rather than only one variable $r,$ and has a
slope discontinuity at $r=r^{\prime}$. For a simple exponential potential an
accuracy of $7$ significant figures is achieved, with the number $N$ of
Chebyshev support points in each partition equal to $17$. For a potential with
a large repulsive core, such as the potential between two $He$ atoms, the
accuracy decreases to $4$ significant figures, but is restored to $7$ if $N$
is increased to $65.$

\end{abstract}
\date[11--28-2007]{}
\startpage{1}
\maketitle

\section{Introduction}

The\ two-body $T-$matrix is most often used for the calculation of two-body
scattering phase shifts by means of integrals, thereby avoiding the need to
match the wave function to known asymptotic functions at large distances. For
two-body scattering calculations a representation of the $T-$matrix in
momentum space is most convenient because this space is most closely related
to the ingoing and outgoing momenta of the scattering process. However, for
other applications it is convenient to formulate the $T-$matrix, and its
respective integral equation, in configuration space. One such application
occurs for the solution in configuration space of the three-body Faddeev
integral equations, which require as input the two-body scattering matrices
for each of the three arrangements \cite{GLOECKLE}.

The function $T(E;r,r^{\prime})$ satisfies the integral equation%
\begin{equation}
T(E;r,r^{\prime})=V(r)\ \delta(r-r^{\prime})+V(r)\int_{0}^{\infty}%
\mathcal{G}_{0}(E;r,\bar{r})\ T(E;\bar{r},r^{\prime})\ d\bar{r} \label{LST}%
\end{equation}
Here $\mathcal{G}_{0}$ is the undistorted (free) two-body Green's function
$\mathcal{G}_{0}=(E+i\varepsilon-H_{0})^{-1},$ that, for a particular partial
wave in configuration space with angular momentum $\ell=0$ is given by the
well known expression%
\begin{equation}
\mathcal{G}_{0}(E;r,\bar{r})=-\frac{1}{k}F(E;r_{<})\ G(E;r_{>}) \label{G0}%
\end{equation}
where $F(E;r)=\sin(kr);G(E;r)=\cos(kr)$ or $\exp(ikr).$ In the case of
\ $\ell\neq0$, $F$ and $G$ are replaced by the corresponding Bessel-Ricatti
functions. Here $k$ is the wave number in units of inverse length, and $E$ and
$V(r)$ are the energy and the local two-body potential, in units of inverse
length squared, respectively. In these units $k=(E)^{1/2}$ (for negative
energies $\kappa=(-E)^{1/2}).$

The purpose of this paper is to present an algorithm for the solution of Eq.
(\ref{LST}) for the two-variable function $T(E;r,r^{\prime})$ based on the use
of spectral expansions in terms of Chebyshev polynomials \cite{IEM},
\cite{DELOFF}. This method, denoted as $IEM$, has been found to be well suited
for the solution of the one-variable Lippmann-Schwinger equation for the wave
function $\psi(r),$both for positive \cite{CISE}, \cite{APPLIC} and negative
\cite{HEHE} energies. One essential feature of the $IEM$ is to divide the
entire radial range into partitions, and expand the unknown function in each
partition in terms of a small number ($\simeq17)$ of basis functions such as
Chebyshev or Legendre polynomials. Partitioning has the merit a) that the
number of mesh points is very economical, since the partitions are
concentrated automatically in those regions where the function varies most
rapidly, b) the accuracy control of the spectral expansions is also
maintained, and c) the singularity in the driving term, described below, can
be easily handled by placing the boundary between two partitions at the
position of the singularity. The basic reason that partitioning can still be
maintained for the calculation of $T$ is that the integral in Eq. (\ref{LST})
is over one variable only, i.e., the second variable is introduced only
parametrically through the driving term $V(r)\ \delta(r-r^{\prime}).$

The properties of $T$ in configuration space are little known, mainly because
$T$ has usually been studied in the momentum representation. However, the
potential between atoms are generally given in configuration space, and the
presence of large repulsive cores as well as long range parts of the
potential, that become important at ultra-cold energies for atom-atom
scattering, can cause difficulties in the momentum representation. It is hoped
that a good understanding of $T$ in configuration space can provide a simpler
way for overcoming these difficulties. Another important motivation is that
the solution of the three-body Faddeev integral equations in momentum space
\cite{gloeckle96}, \cite{golak05} achieves an accuracy of\ three to four
significant figures \cite{PRIVATE}, which is adequate for nuclear physics
applications, but is inadequate for atomic physics applications. In order to
make use of the $IEM$'s high accuracy for the solution of integral equations
in configuration space, the three-body Faddeev integral equations have been
formulated in configuration space \cite{GLOECKLE}. Since a basic input into
these equations are integrals over the two-body $T-$matrices \cite{IT},
knowledge of the properties of the two-body $T-$matrix in configuration space
provides an important contribution.

\section{The formalism}

An important relation between the $T-$matrix, the potential $V$, and the
scattering wave function $\psi$ is
\begin{equation}
V(r)\ \psi(r)=\int_{0}^{\infty}T(r,\bar{r})\ F(\bar{r})d\bar{r}. \label{PSI_T}%
\end{equation}
Here $\psi$ is the solution of the Schr\"{o}dinger equation
\begin{equation}
-\frac{d^{2}\psi}{dr^{2}}+(V-k^{2})\psi=0 \label{SCHR}%
\end{equation}
or of the corresponding Lippmann- Schwinger (L-S) equation
\begin{equation}
\psi(r)=F(r)+\ \int_{0}^{\infty}\mathcal{G}_{0}(r,r^{\prime})\ V(r^{\prime
})\ \psi(r^{\prime})\ dr^{\prime}. \label{PSI_LS}%
\end{equation}
Symbolically the solution of the L-S equation for $\psi$ can be written as
$\psi=(1-\mathcal{G}_{0}\ V)^{-1}F$, from which one finds $T=V\ (1-\mathcal{G}%
_{0}\ V)^{-1}$. The latter can be rewritten as $T=(1-V\mathcal{G}_{0})^{-1}V,$
from which either Eq. (\ref{LST}) follows, or the alternate form
\begin{equation}
T(E;r,r^{\prime})=V(r)\ \delta(r-r^{\prime})+V(r^{\prime})\int_{0}^{\infty
}T(E;r,\bar{r})\mathcal{G}_{0}(E;\bar{r},r^{\prime})\ d\bar{r}, \label{LST_A}%
\end{equation}
obtained by using the formal identity $(1-V\mathcal{G}_{0})^{-1}%
V=V\ (1-\mathcal{G}_{0}V)^{-1}.$

The function $T(r,r^{\prime})$ is symmetric in $r$, and $r^{\prime}$ as can be
seen by comparing Eqs. (\ref{LST}) and (\ref{LST_A}) and making use of the
symmetry of the Green's function $\mathcal{G}_{0}.$ Even though the $T-$matrix
has a discontinuous derivative at $r=r^{\prime},$ as discussed below, the
integral of the product of $T$ with another function $\Phi$
\begin{equation}
\bar{T}(E;r)=\int_{0}^{\infty}T(E,r,r^{\prime})\ \Phi(r^{\prime})\ dr^{\prime}
\label{TBAR}%
\end{equation}
over one of the two variables is a continuous function of the other variable,
and obeys a continuous one-variable integral equation
\begin{equation}
\bar{T}(E;r)=V(r)\ \Phi(r)+V(r)\int_{0}^{\infty}\ \mathcal{G}_{0}(E;r,\bar
{r})\ \bar{T}(E;\bar{r})\ d\bar{r}, \label{LSR}%
\end{equation}
that can be solved without difficulty \cite{IT} using methods described
earlier \cite{IEM}.

The first step in solving Eqs. (\ref{LST}) or (\ref{LST_A}) is to separate out
the delta function term from $T$ by defining $R(r,r^{\prime})$ according to%
\begin{equation}
T(E;r,r^{\prime})=V(r)\ \delta(r-r^{\prime})+R(E;r,r^{\prime}). \label{R1}%
\end{equation}
As a result of Eq. (\ref{LST}) the L-S equation for $R$ becomes
\begin{equation}
R(E;r,r^{\prime})=V(r)\int_{0}^{\infty}\mathcal{G}_{0}(E;r,\bar{r}%
)\ R(E;\bar{r},r^{\prime})\ d\bar{r}+V(r)\mathcal{G}_{0}(E;r,r^{\prime
})V(r^{\prime}), \label{LS_R}%
\end{equation}
which is the object of the present numerical evaluation. One physical
interpretation of $R(E;r,r^{\prime})$ is that its integral over $F$ represents
all second and higher order Born terms of the iterative solution of the L-S
Eq. (\ref{PSI_LS}) for the wave function
\begin{align}
\psi(r)  &  =F(r)+\int_{0}^{\infty}\mathcal{G}_{0}(E;r,\bar{r})\ V(\bar
{r})F(\bar{r})\ d\bar{r}\nonumber\\
&  +\int_{0}^{\infty}\mathcal{G}_{0}(E;r,\bar{r})\left[  \int_{0}^{\infty
}R(E;\bar{r},r^{\prime})\ F(r^{\prime})dr^{\prime}\right]  d\bar{r}.
\label{PSI_BA}%
\end{align}
This equation follows by inserting (\ref{LST}) into the right hand side of Eq.
(\ref{PSI_T}) and cancelling the potential $V(r).$ The second term involving
$R$ in Eq. (\ref{PSI_BA}) is likely to cancel the first two terms in the
repulsive core region of $V$ since $\psi$ is small in this region, hence
integrals over $R$ alone are generally to be avoided.

\section{The Spectral Method}

An extension of the conventional spectral expansion method $IEM$ \cite{IEM}
for solving Eq. (\ref{LS_R}) is described in this section. For simplicity the
energy variable is suppressed in what follows, and the driving term
$V(r)\mathcal{G}_{0}(E;r,r^{\prime})V(r^{\prime})$ is denoted as
$D(E;r,r^{\prime}).$ In order to simplify the notation, it is useful to define
the functions $\mathfrak{F}$ and $\mathfrak{G}$ as
\begin{equation}
\mathfrak{F}(E;r)=F(E;r)V\left(  r\right)  \label{CURLYF}%
\end{equation}%
\begin{equation}
\mathfrak{G}(E;r)=G(E;r)V\left(  r\right)  \label{CURLYG}%
\end{equation}
in terms of which%
\begin{equation}
D(E;r,r^{\prime})=-\frac{1}{k}\mathfrak{F}(E;r_{<})\ \mathfrak{G}(E;r_{>}).
\label{D}%
\end{equation}
Since $D$ has a discontinuous derivative at $r=r^{\prime},$ the solution of
Eq. (\ref{LS_R}) also acquires the same type of discontinuity.

Initially, without regard of the position of $r^{\prime}$, the radial interval
is truncated at an upper limit $0\leq r\leq r_{\max}$ , and that region is
divided into $M$ partitions. The size of each partition $i,$ and the
corresponding number $M$ is determined automatically by the accuracy criterion
according to the properties of Chebyshev expansions.\cite{DELOFF},
(\cite{CISE}), and for this reason these partitions are denoted as Chebyshev
partitions. The lower and upper limits of each partition are denoted by
$t_{i-1}$ and $t_{i}$, respectively,%
\begin{equation}
t_{i-1}\leq r_{i}<t_{i},
\end{equation}
and if $r$ is contained in partition $i$ it is denoted as $r_{i}.$\ A second
mesh of $N_{rp}$ equidistant points $r^{\prime}$ is set up in the radial
interval [$0,$ $r_{\max}$] at a mesh distance of $h_{rp}$. For a point
$r^{\prime}$ on this mesh that falls into a particular Chebyshev partition
$j$, this\ partition is further divided into two, with $r^{\prime}$ located at
the border between the two sub-partitions. The left\ (right) sub-partition is
denoted as $j_{L}$ $(j_{R})$.

In what follows the energy is no longer written explicitly, and the angular
momentum is assumed to be zero. The integral operator $\mathcal{G}_{0}$
restricted to partition $i$ is denoted as $\mathcal{G}_{i}$ and is defined as%
\begin{equation}
\mathcal{G}_{i}\xi\equiv-\frac{1}{k}G(r_{i})\int_{t_{i-1}}^{t_{i}}F(\bar
{r})\xi(\bar{r})\ d\bar{r}-\frac{1}{k}F(r_{i})\int_{r_{i}}^{t_{i}}G(\bar
{r})\xi(\bar{r})\ d\bar{r}, \label{Gi}%
\end{equation}
where $\xi$ is an arbitrary function being operated upon by $\mathcal{G}_{i}%
.$\ Inserting Eq. (\ref{R1}) into Eq.(\ref{LS_R}) one obtains%
\begin{align}
R_{i,j}-V(r_{i})\mathcal{G}_{i}\ R_{i,j}  &  =D(r_{i},r_{j}^{\prime
})+\mathfrak{G}(r_{i})\int_{0}^{t_{i-1}}(-)\frac{1}{k}F(\bar{r})\ R(\bar
{r},r_{j}^{\prime})d\bar{r}\nonumber\\
&  +\mathfrak{F}(r_{i})\int_{t_{i}}^{r_{\max}}(-)\frac{1}{k}G(\bar{r}%
)\ R(\bar{r},r_{j}^{\prime})d\bar{r}. \label{LS_RIJ}%
\end{align}
where the left hand side of Eq. (\ref{LS_RIJ}) denotes $R(r_{i},r_{j}^{\prime
})-V(r_{i})\int_{t_{i-1}}^{t_{i}}\mathcal{G}_{0}(r,\bar{r})\ R(\bar
{r},r^{\prime})\ d\bar{r}.$ The integrals in Eq.(\ref{LS_RIJ}) do not depend
on $r$, and are denoted as
\begin{equation}
A_{i}(r_{j}^{\prime})=-\frac{1}{k}\int_{t_{i}}^{r_{\max}}G(\bar{r})\ R(\bar
{r},r_{j}^{\prime})d\bar{r} \label{AI}%
\end{equation}%
\begin{equation}
B_{i}(r_{j}^{\prime})=-\frac{1}{k}\int_{0}^{t_{i-1}}F(\bar{r})\ R(\bar
{r},r_{j}^{\prime})d\bar{r}. \label{BI}%
\end{equation}
In view of the linearity of the operator $(1-V\mathcal{G}_{i})$ on the left
hand side of Eq. \ref{LS_RIJ}, the solution of Eq. \ref{LS_RIJ} can be written
as \cite{IEM} a linear combination of three functions $Y,\ Z,$ and $S$
\begin{equation}
R\left(  r_{i},r_{j}^{\prime}\right)  =A_{i}(r_{j}^{\prime})\ Y_{i}%
(r_{i})+B_{i}(r_{j}^{\prime})\ Z_{i}(r_{i})+S_{i,j}\left(  r_{i},r_{j}%
^{\prime}\right)  \label{RIJ}%
\end{equation}
that are the solutions of
\begin{equation}
\left[  1-V(r_{i})\mathcal{G}_{i}\right]  \ Y_{i}=\mathfrak{F}(r_{i})
\label{YI}%
\end{equation}%
\begin{equation}
\left[  1-V(r_{i})\mathcal{G}_{i}\right]  \ Z_{i}=\mathfrak{G}(r_{i})
\label{ZI}%
\end{equation}%
\begin{equation}
\left[  1-V(r_{i})\mathcal{G}_{i}\right]  \ S_{ij}=D(r_{i},r_{j}^{\prime})
\label{SIJ_EQ}%
\end{equation}
in each interval $i$. Up to here the procedure is identical to the $IEM$ for
the solution of two-body L-S equations. A new feature is the appearance of the
third function $S_{ij}$, that, in view of Eq.(\ref{D}) is given by
\begin{align}
S_{ij}(r_{i},r_{j}^{\prime})  &  =-\frac{1}{k}Y_{i}(r_{i})\ \mathfrak{G}%
(r_{j}^{\prime})~~~i<j\label{SIJ}\\
S_{ij}(r_{i},r_{j}^{\prime})  &  =-\frac{1}{k}Z_{i}(r_{i})\ \mathfrak{F}%
(r_{j}^{\prime})~~~i>j. \label{SJI}%
\end{align}
For the partition $j$ that subsequently was sub-divided into the left and
right sub-partitions $j_{L}$ and $j_{R}$, with $r_{j}^{\prime}$ at the border
between the two partitions, Eqs. (\ref{SIJ}) and (\ref{SJI}) remain valid for
each of the sub-partitions, provided that $i$ or $j$ are replaced by the
appropriate value $j_{L}$ or $j_{R}.$

The calculations of the coefficients $A$ and $B$ is as follows. Inserting
Eq.(\ref{RIJ}) into Eqs.(\ref{AI}) and (\ref{BI}) one obtains for $i=1,2,..M$%
\begin{equation}
A_{i}(r_{j}^{\prime})=\sum_{h=i+1}^{M}\ v_{h}(r_{j}^{\prime}) \label{Ais}%
\end{equation}%
\begin{equation}
B_{i}(r_{j}^{\prime})=\sum_{h=1}^{i-1}\ u_{h}(r_{j}^{\prime}), \label{Bis}%
\end{equation}
where%
\begin{align}
v_{h}(r_{j}^{\prime})  &  =-A_{h}(r_{j}^{\prime})\ \langle GY\rangle_{h}%
-B_{h}(r_{j}^{\prime})\ \langle GZ\rangle_{h}-\langle GS\rangle_{h,j}%
\label{vs}\\
u_{h}(r_{j}^{\prime})  &  =-A_{h}(r_{j}^{\prime})\ \langle FY\rangle_{h}%
-B_{h}(r_{j}^{\prime})\ FZ\rangle_{h}-\langle FS\rangle_{h,j}. \label{us}%
\end{align}
In the above,
\begin{equation}
\left\langle \eta\xi\right\rangle _{h}=\int_{t_{h-1}}^{t_{h}}\frac{1}{k}%
\eta(\bar{r})\xi_{h}(\bar{r})\ d\bar{r}, \label{GX}%
\end{equation}
where $\eta$ is either $G$ or $F$, and $\xi_{h}$ is either $Y_{h}$ or $Z_{h},$
and%
\begin{equation}
\left\langle \eta S\right\rangle _{h,j}=\int_{t_{h-1}}^{t_{h}}\frac{1}{k}%
\eta(\bar{r})S_{h,j}(\bar{r},r_{j}^{\prime})\ d\bar{r} \label{etaS}%
\end{equation}
From Eqs. (\ref{Ais}) and (\ref{Bis}) it follows that $B_{1}=A_{M}=0$. In view
of Eqs. (\ref{SIJ}) and (\ref{SJI}) one obtains for the inhomogeneous overlap
terms%
\begin{equation}
\left\langle \eta S\right\rangle _{h,j}=-\frac{1}{k}\left\langle \eta
Y\right\rangle _{h}\ \mathfrak{G}(x_{j})~~~h<j \label{OSIJ}%
\end{equation}
and%
\begin{equation}
\left\langle \eta S\right\rangle _{h,j}=-\frac{1}{k}\left\langle \eta
Z\right\rangle _{h}\ \mathfrak{F}(x_{j})~~~h>j. \label{OSJI}%
\end{equation}
Here $h$ stands for any of the partitions $i=1,2,..M,$ with the exception than
when $h=j$, $h$ is replaced by either $j_{L}$ or $j_{R}.$ The equations above
still apply, since $r_{j}^{\prime}$ is located either on the right or left
border of the partition, respectively, as is explained in more detail below.

A recursion relation between the coefficients $A_{h}$ and $B_{h}$ and those
from a neighboring partition $h\pm1$ can be obtained \cite{CISE} by writing
Eqs. Eqs. (\ref{Ais}) and (\ref{Bis}) in the form%
\begin{align}
A_{h+1}-A_{h}  &  =-v_{h+1}\label{REC_A}\\
B_{h+1}-B_{h}  &  =u_{h} \label{REC_B}%
\end{align}
According to Eqs. (\ref{vs}) and (\ref{us}) the functions $v_{h}$ and $u_{h}$
contain the coefficients $A_{h}$ and $B_{h},$ and hence, after some algebra
one obtains, for $i+1$ and $i$ both different from $j$ \cite{CISE}
\begin{equation}
\Omega_{i+1}\left(
\begin{array}
[c]{c}%
A\\
B
\end{array}
\right)  _{i+1}=\Gamma_{i}\left(
\begin{array}
[c]{c}%
A\\
B
\end{array}
\right)  _{i}+\left(
\begin{array}
[c]{c}%
-\left\langle FS\right\rangle _{i,j}\\
\left\langle GS\right\rangle _{i+1,j}%
\end{array}
\right)  ;i=1,2,M-1, \label{AB}%
\end{equation}
where%
\begin{equation}
\Omega_{i+1}=\left(
\begin{array}
[c]{cc}%
0 & 1\\
1-\left\langle GY\right\rangle _{i+1} & -\left\langle GZ\right\rangle _{i+1}%
\end{array}
\right)  \label{OMEGA}%
\end{equation}
and%
\begin{equation}
\Gamma_{i}=\left(
\begin{array}
[c]{cc}%
-\left\langle FY\right\rangle _{i} & 1-\left\langle FZ\right\rangle _{i}\\
1 & 0
\end{array}
\right)  . \label{GAMMA}%
\end{equation}
The matrices $\Omega$ and $\Gamma$ have already been defined in Ref.
\cite{CISE}, while the terms in Eq. (\ref{AB}) involving\ the functions $S$
are new

The method for obtaining the coefficients $A_{i}$ and $B_{i}$ as a function of
$r_{j}^{\prime}$ consists in propagating the coefficients forward from
partition $i=1$ to partition $j_{L}$, and backward, from partition $M$ to
$j_{L}.$ By equating the two results in partition $j_{L},$ all the
coefficients can be determined as a function of $r^{\prime}$. These forward
and backward propagations to the point $r_{j}^{\prime}$ bear some similarity
to the method described in Ref. \cite{HEHE} for the calculation of energy
eigenvalues of the Schr\"{o}dinger equation. Further details for the
calculation of the coefficients $A_{i}$ and $B_{i}$ are given in the Appendix
1. For $i\neq j$ the result is\ a semi-separable expression of the form%
\begin{equation}
R\left(  r_{i},r_{j}^{\prime}\right)  =\left[  A_{i}(r_{j}^{\prime
})-\mathfrak{G}(r_{j}^{\prime})/k\right]  \ Y_{i}(r_{i})+B_{i}(r_{j}^{\prime
})\ Z_{i}(r_{i}),~~i<j \label{RSSIJ}%
\end{equation}%
\begin{equation}
R\left(  r_{i},r_{j}^{\prime}\right)  =A_{i}(r_{j}^{\prime})\ Y_{i}%
(r_{i})+\left[  B_{i}(r_{j}^{\prime})-\mathfrak{F}(r_{j}^{\prime})/k\right]
\ Z_{i}(r_{i}),~~i>j. \label{RSSJI}%
\end{equation}

Integrals of $R(r,r^{\prime})$ over a known function $\Phi(r),$ as defined in
Eq. (\ref{TBAR}), can be obtained by first calculating $R$ and then performing
the required integral, as is done below in order to check the accuracy of the
results for $R$. However, a simpler and more accurate way is to solve the L-S
equations (\ref{LSR}) directly for these integrals. The present code can be
modified for this purpose by replacing the driving term $D$ in Eq. (\ref{D})
by the continuous function $V(r)\ \Phi(r)$ (no right or left sub-partitioning
being required), calculating the function $S$ in each partition according to
Eq. (\ref{SIJ_EQ}), choosing an appropriate point $r^{\prime}$ at the boundary
between two Chebyshev partitions, performing the forward and back ward
recursion relations for the quantities $P,p$ and $Q,q$, respectively, and
finally solving Eqs. (\ref{A1BM}) for the coefficients $A_{1}$ and $B_{M}.$
This operation has to be performed once only for the one point $r^{\prime}$
and can be carried out by the $IEM$ with the same precision as the solution
for the wave function \cite{IEM}, \cite{CISE} or the binding energy
\cite{HEHE}.

If the driving term in Eq. (\ref{D}) were set to zero, the quantities
$S_{i,j}$ in Eq. \ref{SIJ_EQ}) would vanish, and hence Eq. (\ref{LS_R}) would
be of the form%
\begin{equation}
R(E;r,r^{\prime})=V(r)\int_{0}^{\infty}\mathcal{G}_{0}(E;r,\bar{r}%
)\ R(E;\bar{r},r^{\prime})\ d\bar{r}. \label{LS_RH}%
\end{equation}
The solution $R$ of this equation vanishes, unless the determinant of the
discretized form of the operator $1-V(r)\int_{0}^{\infty}d\bar{r}%
\ \mathcal{G}_{0}(E;r,\bar{r})$ is zero. This determinant does vanish for
particular discrete energies which are the bound states of the two-body
system. As the energy $E$ ranges over the negative values which include bound
state energies, poles in the $R$-matrix occur. However, the exploration of
these singularities will be left to a future study.

\subsubsection{Units and Dimensions}

The potential energy $\bar{V}$ and the energy $\bar{E}$\ in the
Schr\"{o}dinger equation are normally given in units of energy, while the
distance is in units of length $(\ell).$ By multiplying all terms in the
Schr\"{o}dinger equation by $2\mu/\hbar^{2},$ where $\mu$\ is the reduced mass
of the two colliding particles, and $\hbar$ is Planck's constant divided by
$2\pi,$ then all terms in the scaled Schr\"{o}dinger equation acquire
dimensions of $\ell^{-2}.$The new potential energy $V$ and the energy $E$ then
are
\begin{equation}
V=Q\bar{V} \label{VBAR}%
\end{equation}%
\begin{equation}
E=k^{2}=Q\bar{E} \label{EBAR}%
\end{equation}
where $Q$ is a scaling factor.

For nuclear physics applications $\bar{E}$ and $\bar{V}$ are given in $MeV$,
the radial distance in $fm$ and the corresponding value of $Q$ is%
\begin{equation}
Q_{N}=2\mu/\hbar^{2}. \label{QN}%
\end{equation}
The corresponding scaled Schr\"{o}dinger equation is Eq. (\ref{SCHR}), and the
corresponding L-S equation is (\ref{PSI_LS}).The dimension of the Green's
function $\mathcal{G}_{0},$ Eq. (\ref{G0}), is $\ell^{1}$ $($for the nuclear
case $\ell\equiv fm$) and hence the operator $\mathcal{G}_{0}(r,r^{\prime
})V(r^{\prime})dr^{\prime}$ has no dimension. In view of Eq. (\ref{PSI_T}),
and since $F$ and $\psi$ have no dimension, the $T-$Matrix has dimension
$\ell^{-3}$. This dimension is compatible with Eq. (\ref{R1}), since the delta
function has dimension $\ell^{-1}.$ The dimension of $R$ is also $\ell^{-3},$
which is compatible with Eq. (\ref{LS_R}). The dimension of $A$ and $B$, Eqs.
(\ref{AI}) and (\ref{BI}), is $\ell^{-1}$, the dimension of $Y$ and $Z,$ Eqs.
(\ref{YI})-(\ref{ZI}), is $\ell^{-2},$ and that of $S,$ (\ref{SIJ_EQ}) is
$\ell^{-3}.$ The quantities $\left\langle \eta\xi\right\rangle ,$ Eq.
(\ref{GX}), and $\left\langle \eta S\right\rangle $ Eq. (\ref{etaS}), have
dimensions $\ell^{0}$ and $\ell^{-1}$, respectively. The quantities
$\mathfrak{G}$ or $\mathfrak{F}$ have dimensions $\ell^{-2},$ and hence Eqs.
(\ref{RSSIJ}) and (\ref{RSSJI}) are dimensionally self-consistent, each term
having the dimension $\ell^{-3}.$

For atomic physics applications $\bar{E}$ is given in atomic energy units
$2\mathbb{R}$, ($\mathbb{R\simeq}13.606\ eV$) and the distance $r$ in units of
the Bohr radius $a_{0},$ in which case%
\begin{equation}
Q_{A}=\frac{2\mu}{\hbar^{2}}a_{0}^{2}\times2\mathbb{R}=\frac{2\mu}{m_{e}},
\label{QA}%
\end{equation}
where $m_{e}$ is the mass of the electron. In this case $\bar{E}$ and $r$, and
similarly the scaled quantities $V$ and $E$, Eqs. (\ref{VBAR}) and
(\ref{EBAR}), respectively, have no dimension and hence all the quantities
described above for the nuclear case have no dimensions either.

\section{Numerical Examples}

Two different potentials in Eq. (\ref{SCHR}) are used for the numerical
examples presented here. One is a simple exponential potential and the other
is a potential describing the interaction between two Helium atoms, based on
the potential $TTY$ \cite{TTY}, \cite{Franco}. For the He-He case a soft
repulsive core is applied for a distance less than $r_{cut}=4.5.$ The
numerical $IEM$ calculations are carried out from $r=0$ to $r=r_{\max}$, so as
to include the effect of the core accurately. The value of the accuracy
parameter is $tol=10^{-8}$, and the wave number is $k=1.5$. The potential and
the wave number are in units of inverse length squared, and inverse length, respectively.

\subsection{The exponential example.}

\begin{figure}
[ptb]
\begin{center}
\includegraphics[
height=3.5362in,
width=4.7003in
]%
{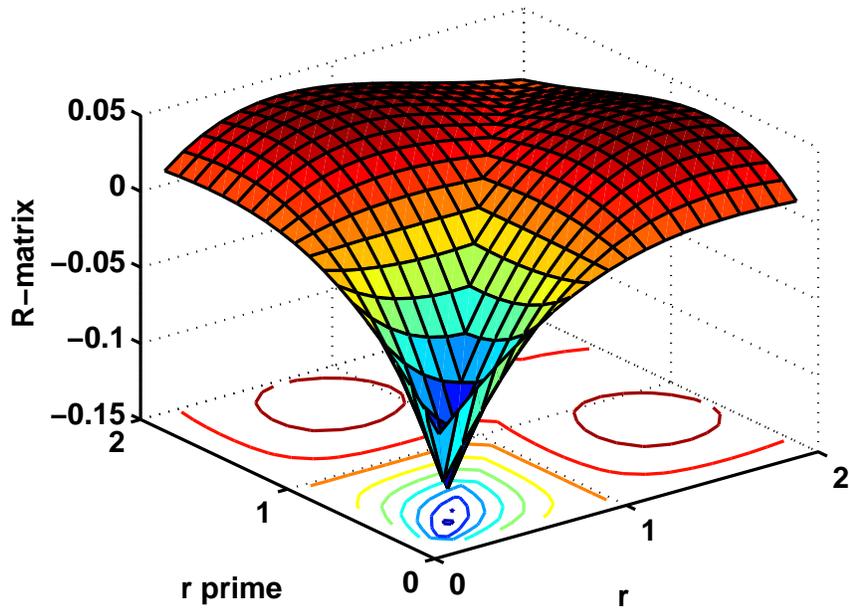}%
\caption{The $R$-matrix for the exponential potential $V=\exp(-r)$. The wave
number $k$ is$1.5.$ The discontinuity in the derivative, for $r=r^{\prime}$ is
clearly visible.}%
\label{R_exp}%
\end{center}
\end{figure}

\begin{figure}
[ptb]
\begin{center}
\includegraphics[
height=3.5362in,
width=4.7003in
]%
{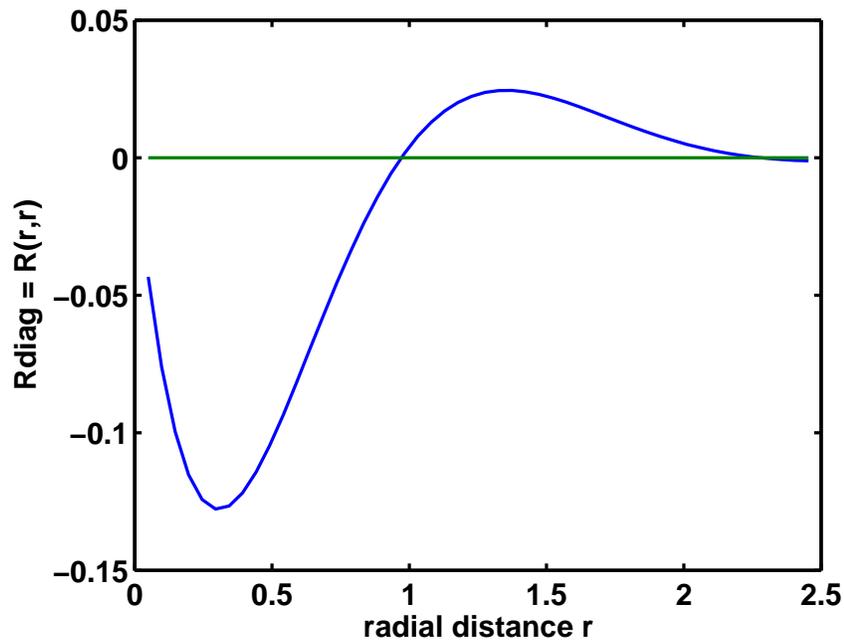}%
\caption{The diagonal part of the $R-$matrix, shown in Fig. (\ref{R_exp}).}%
\label{R_diag_exp}%
\end{center}
\end{figure}

The potential in Eq. (\ref{SCHR}) is $V(r)=\pm\exp(-r)$.\ The $R$-matrix for
the repulsive case is shown in Fig (\ref{R_exp}), and the
behavior of the diagonal part of $R$ is shown in Fig. (\ref{R_diag_exp}). The
derivative discontinuity of $R(r,r^{\prime})$ at $r=r^{\prime}$ is clearly
visible in Fig. (\ref{R_exp}), and\ it is also seen that $R$ is large where
the potential is large. These features are also present for the $He-He$ case,
described below.

\subsection{The He-He case.\bigskip}

\begin{figure}
[ptb]
\begin{center}
\includegraphics[
height=3.5362in,
width=4.7003in
]%
{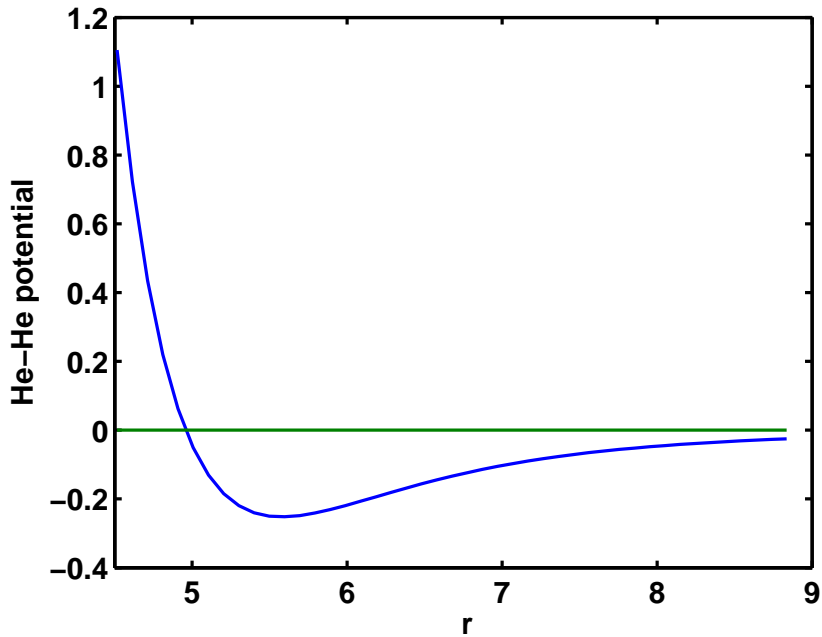}%
\caption{The $He-He$ atom interaction, based on the potential TTY \cite{TTY}.
The units for the distance are $a_{0}$ (TheBohr radius) and $(a_{0})^{-2}%
$\ for the potential. For distances less than $4.5\ a_{0}$ a repulsive soft
core is matched to this potential, as described in the text. At $r\simeq0$ its
value is $\simeq\ 150$ $(a_{0})^{-2}$\ . The $He-He$ dimer is very weakly
bound, with a binding energy of $2.58\times10^{-5}\ (a_{0})^{-2}$, or
$\simeq10^{-7}eV$. Upon division by the factor $7296.3$ this potential is
transformed to atomic energy units, as described in Ref.. \cite{HEHE}.}%
\label{POT}%
\end{center}
\end{figure}

For the $He-He$ case the potential is shown in Fig. (\ref{POT})

It has a strong repulsive core, that is not adequately represented by the
formula described in Refs. \cite{TTY},\cite{Franco}. Since the result for the
$R-$matrix depends significantly on the nature of the repulsive core, an
"artificial" repulsive soft core was used to replace the repulsive core of
$TTY$ for distances $0<r\leq r_{cut}$. This core is of the form $a+br+cr^{2}$,
with the coefficients $a,b,$and $c$ determined such that the repulsive core is
equal to the value, the first, and second derivatives of $TTY$ at $r=r_{cut}$.
This form of the core is less unphysical than the "cut-off" core used in our
previous numerical calculations of the binding energy of the $He-He$ dimer
\cite{HEHE}, for which $V$ was replaced by a constant value $V(r_{cut})$ for
$0<r\leq r_{cut}$.

\begin{figure}
[ptb]
\begin{center}
\includegraphics[
height=3.5362in,
width=4.7003in
]%
{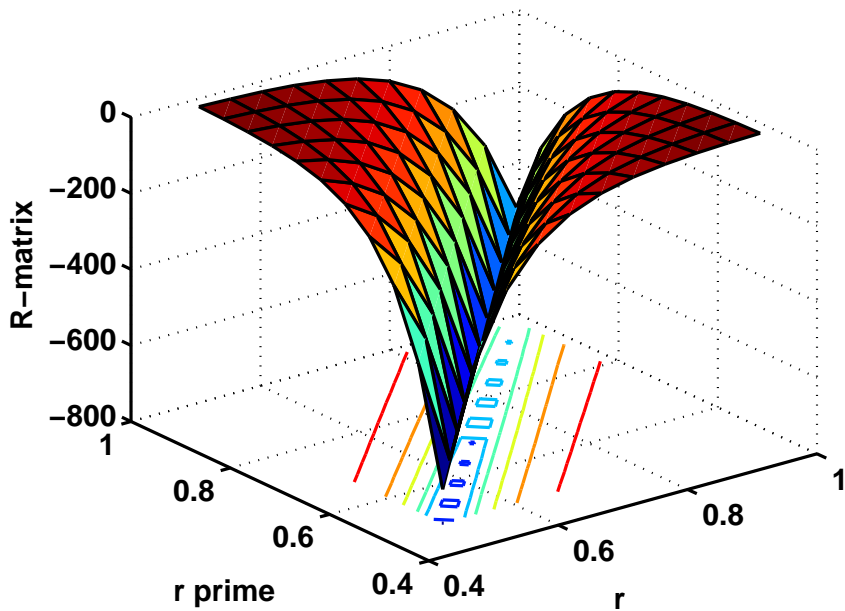}%
\caption{The $R-$matrix for the $He-He$ potential at distances well inside of
the repulsive core, with $r_{cut}=4.5,$ $\ $and $k=1.5.$}%
\label{R_2_He}%
\end{center}
\end{figure}

\begin{figure}
[ptb]
\begin{center}
\includegraphics[
height=3.5362in,
width=4.7003in
]%
{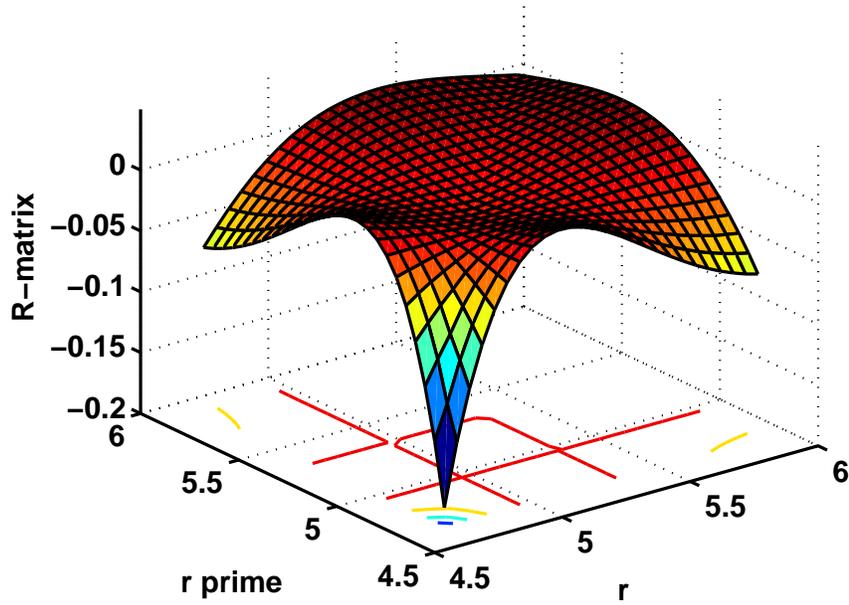}%
\caption{The He-He $R-$matrix, same as in Fig. (\ref{R_2_He}), but at larger
distances, straddling the end of the repulsive core and the attractive
potential valley.}%
\label{R_4-6_He}%
\end{center}
\end{figure}

\begin{figure}
[ptb]
\begin{center}
\includegraphics[
height=3.5362in,
width=4.7003in
]%
{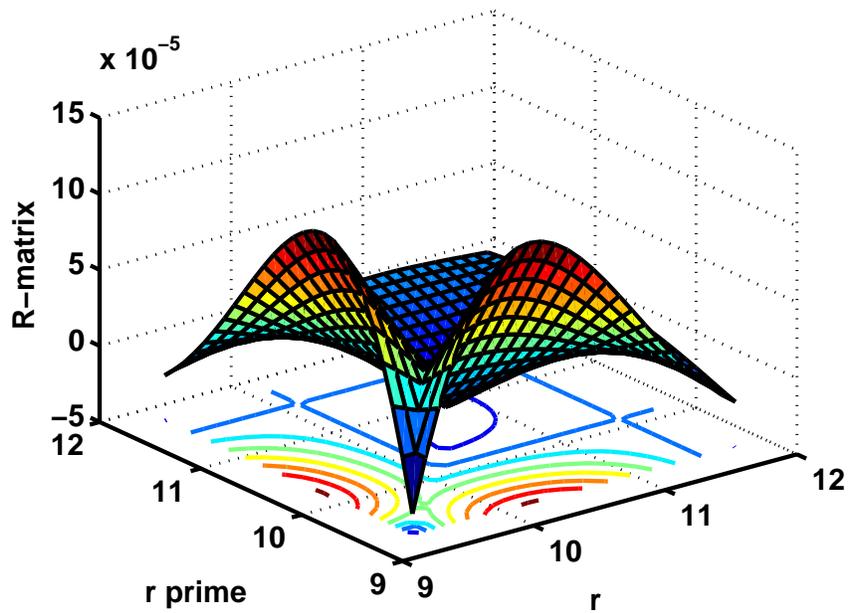}%
\caption{The He-He $R-$matrix, same as in Fig. (\ref{R_2_He}), at distances
between $9$ and $12$ $a_{0}$ where the potential is of the $r^{-6}$ form. The
value of $R$ is of the order of $10^{-5}$ ($a_{0})^{-3}.$}%
\label{RHeHe_9_12}%
\end{center}
\end{figure}

The $R-$matrix at small distances is shown in Fig. (\ref{R_2_He}). The values
of $R$ are very large in the region or the large repulsive core, but they
become smaller at larger distances, Figs. (\ref{R_4-6_He}) and
(\ref{RHeHe_9_12}), for which the potential becomes progressively smaller.

\begin{figure}
[ptb]
\begin{center}
\includegraphics[
height=3.5362in,
width=4.7003in
]%
{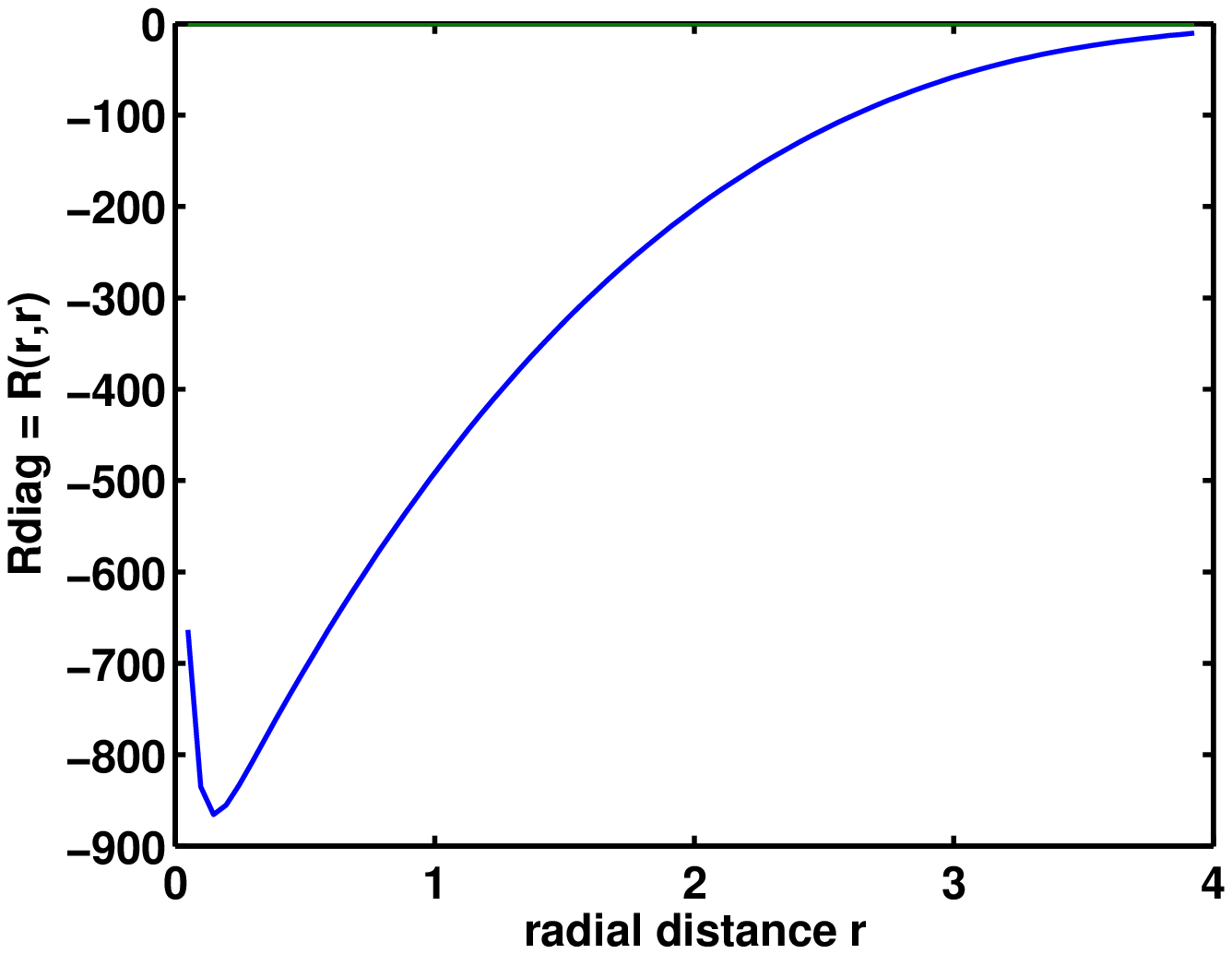}%
\caption{The diagonal part of the $He-He$ $R$-matrix, shown in Fig.
(\ref{R_2_He}). at small distances}%
\label{R_Diag_He}%
\end{center}
\end{figure}

\begin{figure}
[ptb]
\begin{center}
\includegraphics[
height=3.5362in,
width=4.7003in
]%
{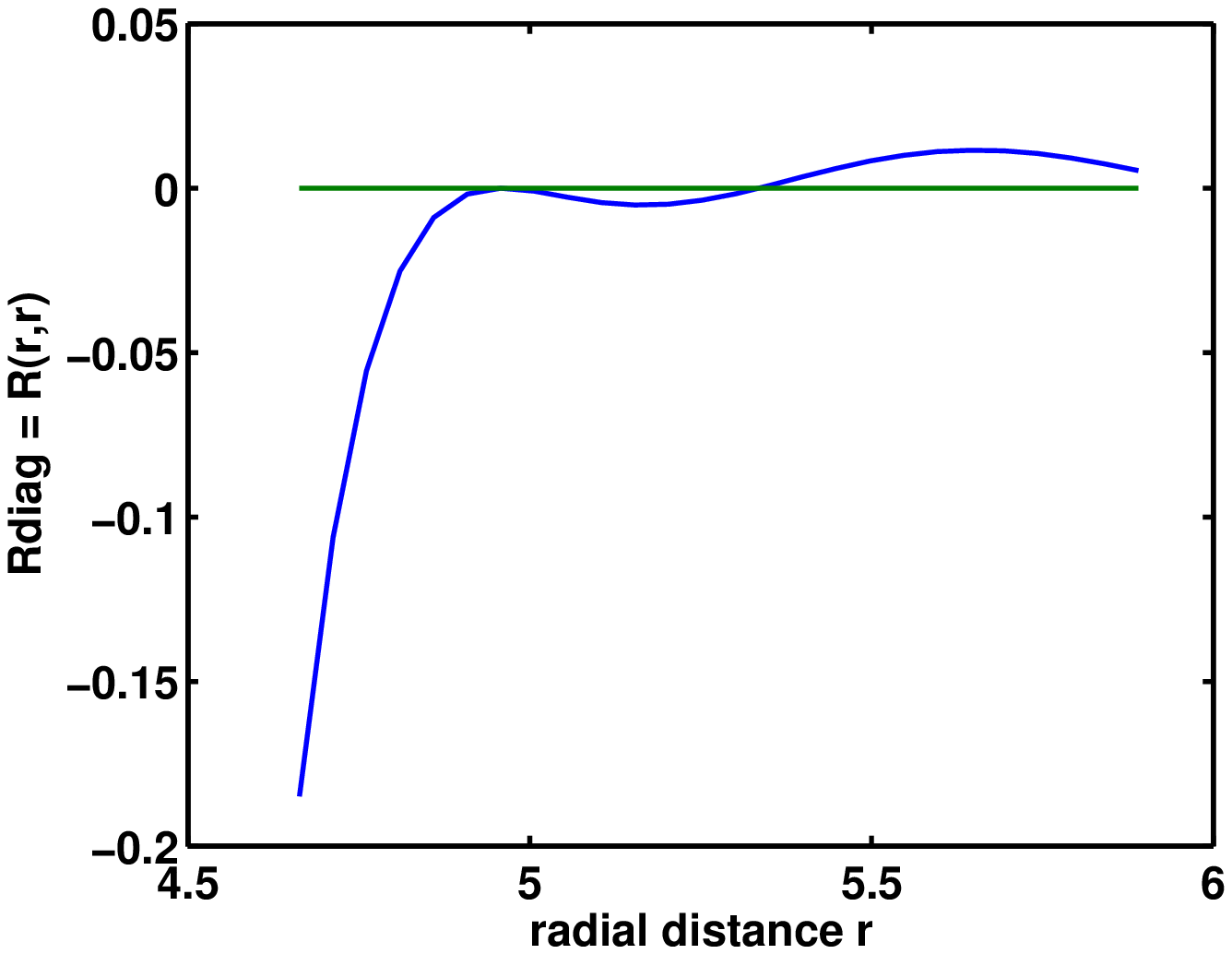}%
\caption{Same as Fig. (\ref{R_Diag_He}) for large distances.}%
\label{R_Diag_4-6_He}%
\end{center}
\end{figure}

The diagonal parts of the $He-He$ $R-$matrix are shown in Figs.
(\ref{R_Diag_He}) and (\ref{R_Diag_4-6_He}), for short and large distances,
respectively.%

\subsection{Accuracy Tests.}

The accuracy of the functions $Y$ and $Z$ in each partition, Eqs. (\ref{YI})
and (\ref{ZI}), determined by the size of the partitions, is better than
$tol=10^{-8}$. In order to obtain a measure of the loss of accuracy that takes
place in the subsequent steps of the calculation, the following two tests are
performed: one that checks the symmetry of the $R-$matrix, and the other that
performs the numerical integral
\[
\mathfrak{I(r}^{\prime})=\int_{0}^{r_{\max}}R(r,r^{\prime})\times F(r)dr
\]
which, according to Eq. (\ref{PSI_T}), should equal $[\psi(r^{\prime
})-F(r^{\prime})]\ V(r^{\prime}).$
\begin{equation}
\mathfrak{I}(r^{\prime})=?~[\psi(r^{\prime})-F(r^{\prime})]\ V(r^{\prime}),
\label{I_CHECK}%
\end{equation}
In Table \ref{TABLE1} the values of the two sides of the above equation are
compared in order to check the accuracy of their agreement with each other.
\begin{table}[tbp] \centering
\begin{tabular}
[c]{|l|l|l||l|l|}\hline
$n$ & $r=\frac{n\times\pi}{16}$ &  & $~~~~~~~Exp.~pot\prime l$ &
$~~~~~He-He~pot^{\prime}l$\\\hline\cline{4-5}%
$10$ & $\simeq1.96$ & $L$ & $\mathbf{3.85431605}$ $\mathbf{[}-2]$ &
$\mathbf{-1.079593140\ [}1]$\\\hline
&  & $R$ & $3.85431605\ [-2]$ & $-1.079593140\ [1]$\\\hline
$15$ & $\simeq2.95$ & $L$ & $\mathbf{4.052449}38\ [-3]$ &
$\mathbf{2.34862829\ [}1]$\\\hline
&  & $R$ & $4.05244931\ [-3]$ & $2.34862833\ [1]$\\\hline
$20$ & $\simeq3.93$ & $L$ & $\mathbf{-5.633010}764\ [-3]$ &
$\mathbf{4.12156423}$\\\hline
&  & $R$ & $-5.633010723\ [-3]$ & $4.12156414$\\\hline
$25$ & $\simeq4.91$ & $L$ & $-\mathbf{1.0735780}42\ [-3]$ &
$\mathbf{1.4144533\ [}-2]$\\\hline
&  & $R$ & $-1.073578029\ [-3]$ & $1.4144530\ [-2]$\\\hline
30 & $\simeq5.89$ & $L$ & $\mathbf{7.192930}11\ [-4]$ & $\mathbf{8.2459426\ [}%
-2]$\\\hline
&  & $R$ & $7.19293028\ [-4]$ & $8.2459429\ [-2]$\\\hline
35 & $\simeq6.87$ & $L$ & $\mathbf{2.0552206}7\ [-4]$ & $\mathbf{3.800}%
32458\ [-2]$\\\hline
&  & $R$ & $2.05522071\ [-4]$ & $3.80032395\ [-2]$\\\hline
40 & $\simeq7.85$ & $L$ & $-\mathbf{8.596136}54\ [-5]$ &
$\mathbf{1.9356054\ [}-2]$\\\hline
&  & $R$ & $-8.59613675\ [-5]$ & $1.9356055\ [-2]$\\\hline
\end{tabular}
\caption{Values of the right (R) and left (L) sides of Eq. (\ref{I_CHECK}). Numbers in square brackets denote powers of 10}\label{TABLE1}%
\end{table}
For the exponential case the number of Chebyshev support points in each
partition is $N=17,$ and only four partitions are required to cover the
interval from $r=0$ to $r=25$.

For $He-He$ case, if $N=17,$ the interval from $0$ to $250$ requires $26$
partitions, $11$ of which lie in the region of the repulsive core. The
corresponding accuracy of the integral $\mathfrak{I}(r)$ (not displayed in the
table) is three to four significant figures. As $N$ is increased, the size of
each partition increases, correspondingly the number of partitions decreases.
The accuracy of $R$ increases the smaller the number of partitions, because
the Chebyshev support points are the ones that carry the large change in the
$R-$values, rather than the coefficients $A$ and $B.$. The results for
$\mathfrak{I}(r)$ for $N=65$ are shown in "L" lines of Table (\ref{TABLE1}),
from which one can deduce an accuracy for $R$ of at least $7$ significant
figures. For this case there are only four partitions in the region
$[0\longrightarrow5]$ \ of the repulsive core, and four more partitions that
cover the whole remaining distance $[5\longrightarrow250]$. These accuracy
results for both the exponential and the $He-He$ cases are confirmed by the
degree of symmetry of the $R-$matrix.

Using MATLAB, version $7.0.1.24704(R14)$ on a $2.8$ GHz intel computer, the
total time required for a $He-He$ calculation of $R(r,r^{\prime})$, with
$n=1273$ equidistant meshpoints for each $r$ and $r^{\prime}$, $r_{\max}=250$,
$N=17$, and $tol=10^{-8}$ requires $114$ seconds. (This time scales like
$n^{2})$ \ For the same calculation, but with $N=65$, the total MATLAB time is
$261$ s$.$ The corresponding calculation of the wave function $\psi,$ for
$r_{\max}=1500$ and also $1273$ equidistant meshpoints takes between $1$ and
$2$ seconds.

\bigskip

\section{Summary and conclusions}

In this paper the spectral IEM method for solving the one variable
Lippmann-Schwinger\ integral equation ($L-S$) in configuration space for the
wave function \cite{IEM},\cite{CISE} is extended to the calculation of the two
variable two-body $T-$matrix, also in configuration space. The $T$-matrix is
written as the sum of a delta-function multiplying the (local) potential plus
a reminder, called the $R$-matrix. The $L-S$ equation for the latter is the
object of the calculation. The main complication is that the driving term for
this equation is continuous but has a discontinuous derivative at
$r=r^{\prime}$. That discontinuity also propagates into the $R$-matrix, and
makes the calculation more cumbersome. This difficulty is overcome by
constructing the partitions into which the radial domain is divided in such a
way that the point $r^{\prime}$ is located at the end-point of a partition.
This procedure is repeated for all values of $r^{\prime}$ contained in an
equidistant set of mesh of points. Otherwise the calculation is very similar
to the spectral $IEM$ procedure for the solution of the one variable $L-S$
equation, whose accuracy and reliability was previously demonstrates in
various applications \cite{APPLIC},\cite{HEHE}. The present method complements
a previous investigation of expanding $T$ in terms of a sum over Sturmian
functions \cite{STURM}.

Two numerical examples are described. One, in which the potential is a simple
exponential function of $r$, and the other in which the potential describes
the interaction between two Helium atoms \cite{TTY}. For the exponential case
the high accuracy, which is determined by the inputted tolerance parameter
$tol$, is maintained. For $tol=10^{-8}$, the $R-$matrix is accurate to $7$
significant figures. For the $He-He$ case the accuracy of $R$ is reduced to
between $3$ and $4$ significant figures. This loss of accuracy is attributed
to the presence of the large repulsive core present in the $He-He$
interaction. If the number of Chebyshev support points in each partition is
increased from $17$ to $65,$ then the accuracy of the $R-$matrix for the
$He-He$ case increases to $7$ significant figures.

The calculation of integrals $\bar{T}(r)$ over $T(r,r^{\prime})$ can be
carried \cite{IT}\ out with far less computational effort than the calculation
of $T(r.r^{\prime}),$ since these integrals can be obtained from the solution
of a $L-S$ equation (\ref{LSR}) that is very similar to the $L-S$ equation for
the wave function, whose accuracy was found to be close to $7$ or $8$
significant figures \cite{HEHE}. This is of interest, since the integral
Faddeev equations for the three-body system, formulated in configuration space
\cite{GLOECKLE} have such integrals as important inputs. It is our aim to
investigate the numerical solution of the three-body integral Faddeev
equations in configuration space in a future study, using the results of the
present paper, and that of Ref. \cite{IT}.

\bigskip

{\Large Acknowledgments:} The author is greatly indebted to Professor Walter
Gl\"{o}ckle for stimulating and consistent help with this work, and for
pointing out its underlying importance with regards to the envisaged
three-body calculations. The author is also grateful to Professor\ I.
Koltracht, at UConn, for help with the computational-mathemathical aspects of
this work.

\bigskip

{\LARGE Appendix 1. The calculation of the coefficients A and B.}\bigskip

Implementing the solution of the recursion relations (\ref{AB}) in a
numerically efficient way will be described below. A short summary is as
follows. By implementing the forward recursion relations (\ref{FRI}) described
below, starting with $(A_{1},B_{1})=(1,0)$ one obtains the quantities $P_{i}$
and $p_{i}$, Eq. \ref{ABPI}), in terms of which the corresponding values of
$A_{i}$ and $B_{i}$ for $i<j$, are given by $(A_{i},B_{i})^{T}=A_{1}P_{i}+$
$p_{i}$, Eq.\ (\ref{ABFP}) with $A_{1}$ still unknown (here $T$ means
"transposed"). Similarly, the coefficients $A_{i}$ and $B_{i}$ are propagated
backwards from the last partition $M$ also into $j_{L}$, using Eq.
(\ref{BRI}), starting with $(A_{M},B_{M})=(0,1)$ , Eq.\ (\ref{ABFP}). The
result is given by a combination of the quantities $Q_{i}$ and $q_{i}$, Eq.
\ref{ABQI}). In terms of these quantities the corresponding coefficients
$A_{i}$ and $B_{i}$ for and for $i>j$ are given by $(A_{i},B_{i})^{T}%
=B_{M}Q_{i}+$ $q_{i}$, Eq.\ (\ref{ABFQ}) with $B_{M}$ still unknown. By
equating the coefficients $A_{j_{L}}$ and $B_{j_{L}}$ in the same partition
$j_{L},$ obtained by the two procedures, one obtains two equations for the
coefficients $A_{1}$ and $B_{M}$ in terms of the known "driving" terms $p_{i}$
and $q_{i}$ involving the functions $S_{i,j}.$ These equations\ (\ref{A1BM})
then can be solved for $A_{1}$ and $B_{M},$ but since they depend on the
location of $r_{j}^{\prime}$, the resulting values of $A_{1}$ and $B_{M}$ also
depend on $r_{j}^{\prime}.$ By repeating this procedure for all values of
$r_{j}^{\prime}$ in partition $j$ for all partitions $j$ one obtains $A_{i}$
and $B_{i}$ for all values of $r_{j}^{\prime}.$

In detail the procedure is as follows. In view of Eq. (\ref{AB}) the "forward"
recursion relations are
\begin{equation}
\left(
\begin{array}
[c]{c}%
A^{f}\\
B^{f}%
\end{array}
\right)  _{i+1}=\left(  \Omega_{i+1}\right)  ^{-1}\left\{  \Gamma_{i}\left(
\begin{array}
[c]{c}%
A^{f}\\
B^{f}%
\end{array}
\right)  _{i}+\frac{\mathfrak{G}(r_{j}^{\prime})}{k}\left(
\begin{array}
[c]{c}%
\left\langle FY\right\rangle _{i}\\
-\left\langle GY\right\rangle _{i+1}%
\end{array}
\right)  \right\}  ,\ i=1,2,..j-2. \label{FRI}%
\end{equation}
In the above, when $i=j-1$ then $i+1=j_{L}.$ By repeatedly using Eqs.
(\ref{FRI}) one can express $A_{i}$ and $B_{i}$ in terms of $A_{1}$ and
$B_{1},$ with the result%
\begin{equation}
\left(
\begin{array}
[c]{c}%
A^{f}\\
B^{f}%
\end{array}
\right)  _{i}=P_{i}+p_{i}(j),~i=2,3,..,j-1,j_{L}, \label{ABPI}%
\end{equation}
where the recursion for the $P_{i}$ and $p_{i}$ column vectors for
$i=1,2,..j-1$ are
\begin{equation}
P_{1}=\left(
\begin{array}
[c]{c}%
1\\
0
\end{array}
\right)  ;~~P_{i+1}=\Omega_{i+1}^{-1}\ \Gamma_{i}\ P_{i}, \label{REC_P}%
\end{equation}%
\begin{equation}
p_{1}=\left(
\begin{array}
[c]{c}%
0\\
0
\end{array}
\right)  ;~p_{i+1}(r_{j}^{\prime})=\Omega_{i+1}^{-1}\ \left[  \Gamma
_{i}\ p_{i}(r_{j}^{\prime})+\frac{1}{k}\mathfrak{G}(r_{j}^{\prime})\left(
\begin{array}
[c]{c}%
\left\langle FY\right\rangle _{i}\\
-\left\langle GY\right\rangle _{i+1}%
\end{array}
\right)  \right]  . \label{REC_p}%
\end{equation}
When $i=j-1,$ the above expressions yield $P_{j_{L}}$ and $p_{j_{L}.}$

The "backward" recurrence relations are%
\begin{equation}
\left(
\begin{array}
[c]{c}%
A^{b}\\
B^{b}%
\end{array}
\right)  _{i-1}=\left(  \Gamma_{i-1}\right)  ^{-1}\left\{  \Omega_{i}\left(
\begin{array}
[c]{c}%
A^{b}\\
B^{b}%
\end{array}
\right)  _{i}+\frac{\mathfrak{F}(r_{j}^{\prime})}{k}\left(
\begin{array}
[c]{c}%
-\left\langle FZ\right\rangle _{i-1}\\
\left\langle GZ\right\rangle _{i}%
\end{array}
\right)  \right\}  ,\ i=M,M-1,..,j+1. \label{BRI}%
\end{equation}
When $i=j+1$ then $i-1$ is replaced by $j_{R}.$ The superscripts $f$ and $b$
stand for "forward" and "backward", respectively. By repeatedly using Eq.
(\ref{BRI}) one obtains
\begin{equation}
\left(
\begin{array}
[c]{c}%
A^{b}\\
B^{b}%
\end{array}
\right)  _{i}=Q_{i}+q_{i}(j),~i=M-1,..,j+1,j_{R} \label{ABQI}%
\end{equation}
and the corresponding recursion relations for the $Q_{i}$ and $q_{i}$ column
vectors are
\begin{equation}
Q_{1}=\left(
\begin{array}
[c]{c}%
0\\
1
\end{array}
\right)  ;~~Q_{i-1}=\Gamma_{i-1}^{-1}\ \Omega_{i}\ Q_{i} \label{REC_Q}%
\end{equation}%
\begin{equation}
q_{1}=\left(
\begin{array}
[c]{c}%
0\\
0
\end{array}
\right)  ;~q_{i-1}(r_{j}^{\prime})=\Gamma_{i-1}^{-1}\ \left[  \Omega
_{i}\ q_{i}(r_{j}^{\prime})+\frac{\mathfrak{F}(r_{j}^{\prime})}{k}\left(
\begin{array}
[c]{c}%
-\left\langle FZ\right\rangle _{i-1}\\
\left\langle GZ\right\rangle _{i}%
\end{array}
\right)  \right]  . \label{REC_q}%
\end{equation}
In these expressions $r_{j}^{\prime}$ stands to the left of any partition,
hence Eqs. (\ref{BRI}) apply, and Eqs. (\ref{REC_Q}) are valid for
$i-1=M-2,M-1,..,$ $j_{L}$ while Eqs. (\ref{REC_q}) are valid only for
$i-1=M-2,M-1,..,j_{R}.$ For the backward recursion from $j_{R}$ to $j_{L}$ the
expression for $q_{j_{L}}$ has to be replaced by%
\begin{equation}
q_{j_{L}}(r_{j}^{\prime})=\Gamma_{j_{L}}^{-1}\ \left[  \Omega_{j_{R}%
}\ q_{j_{R}}(r_{j}^{\prime})+\frac{1}{k}\left(
\begin{array}
[c]{c}%
-\mathfrak{G}(r_{j}^{\prime})\left\langle FY\right\rangle _{j_{L}}\\
\mathfrak{F}(r_{j}^{\prime})\left\langle GZ\right\rangle _{j_{R}}%
\end{array}
\right)  \right]  \label{qjL}%
\end{equation}

The vectors $P_{i}$ \ for $i<j$ and $Q_{i}$ for $i>j$ are independent of
$r_{j}^{\prime}$ and the corresponding vectors $p_{i}$ and $q_{i}$ all
dependent on $r_{j}^{\prime}$ via the factor $\mathfrak{G}(r_{j}^{\prime})/k$
and $\mathfrak{F}(r_{j}^{\prime})/k$ respectively, a feature that provides
economy in the numerical calculation. The vectors $(A^{f},B^{f})^{T}$ as given
by Eq. (\ref{ABPI}) are the values of the coefficients $A$ and $B$ when the
values of $(A_{1},B_{1})^{T}$ in the first partition are set to $(1,0)^{T}$
\ and similarly for $(A^{b},B^{b})^{T}$, as given by Eq. (\ref{ABQI}), when
the values of $(A_{M},B_{M})^{T}$ in the last partition $M$ are set to
$(0,1)^{T}.$ The correct values of $(A,B)^{T}$ are obtained by multiplying the
vectors $P_{i}$ by $A_{1}$ and the vectors $Q_{i}$ by $B_{M},$ i.e.,%
\begin{equation}
\left(
\begin{array}
[c]{c}%
A\\
B
\end{array}
\right)  _{i}=A_{1}(r_{j}^{\prime})\ P_{i}+p_{i}(r_{j}^{\prime}%
),~i=2,3,..,j-1,j_{L}, \label{ABFP}%
\end{equation}%
\begin{equation}
\left(
\begin{array}
[c]{c}%
A\\
B
\end{array}
\right)  _{i}=B_{M}(r_{j}^{\prime})\ Q_{i}+q_{i}(r_{j}^{\prime}%
),~i=M-1,..,j+1,j_{R}. \label{ABFQ}%
\end{equation}
In order to determine the values of $A_{1}$ and $B_{M}$ one equates to each
other the values of $(A,B)^{T}$ obtained by Eqs. (\ref{ABFP}) and
(\ref{ABFQ}), respectively, in the partition $j_{L}$ with the result
\begin{equation}
\left(
\begin{tabular}
[c]{ll}%
$P^{(1)}$ & $-Q^{(1)}$\\
$P^{(2)}$ & $-Q^{(2)}$%
\end{tabular}
\ \ \right)  _{j_{L}}\left(
\begin{tabular}
[c]{l}%
$A_{1}$\\
$B_{M}$%
\end{tabular}
\ \ \right)  =q_{j_{L}}-p_{j_{L}}. \label{A1BM}%
\end{equation}

The procedure described above is repeated for every value of $r^{\prime}$ in
partition $j.$ A convenient set of such values, denoted as $r_{j,s^{\prime}%
}^{\prime},$ $s^{\prime}=1,2,..N+1,$ which originate from in the equidistant
mesh of points $r^{\prime}$ and are located in the Chebyshev mesh $j$.

The case when $i=j$ is as follows, but for clarity of notation the subscript
$j,$ is suppressed. Each $L_{s^{\prime}}$ and $R_{s^{\prime}}$ partition has
its own $N+1$ Chebyshev support points by means of which the Chebyshev
expansion coefficients of the functions $Y_{L_{s^{\prime}}}$ and
$Z_{L_{s^{\prime}}},$ as well as $Y_{R_{s^{\prime}}}$ and $Z_{R_{s^{\prime}}}$
are calculated via the $IEM$ in the Left and Right sub-partitions. By means of
these expansion coefficients the values of the functions $Y$ and $Z$, and
hence also the values of these functions at a point $r$ contained in the
equidistant set of mesh-points can be calculated, and hence the functions $R$
can be obtained at the equidistant mesh points of $r$ and $r^{\prime}.$ In a
simplified notation, $R_{j}\left(  s,s^{\prime}\right)  =\left[  A(s^{\prime
})-\mathfrak{G}(s^{\prime})/k\right]  \ Y_{L_{s^{\prime}}}(s)+B(s^{\prime
})\ Z_{L_{s^{\prime}}}(s),$ $s\leq s^{\prime}$, and a similar expression based
on Eq. (\ref{RSSJI}) for $s>s^{\prime}.$

After the cases that the points $r$ are contained in different Chebyshev
partitions than point $r^{\prime}$, are combined with the case where $r$ and
$r^{\prime}$ are contained in the same Chebyshev partition, one obtains the
final result for $R(r_{s},r_{s^{\prime}}^{\prime}).$ The dimension of this
matrix is equal to the number $N_{rp}$ of equidistant mesh points.

\end{document}